\documentclass[12pt]{article}
\pdfoutput=1

\usepackage[bulletsep]{collref}
\usepackage{amssymb,graphicx}
\usepackage[intlimits]{amsmath}
\usepackage{bbm}
\usepackage[small]{subfigure}

\usepackage{MnSymbol}

\makeatletter \@addtoreset{equation}{section} \makeatother

\makeatletter
\let\old@startsection=\@startsection
\let\oldl@section=\l@section
\renewcommand{\@startsection}[6]{\old@startsection{#1}{#2}{#3}{#4}{#5}{#6\mathversion{bold}}}
\renewcommand{\l@section}[2]{\oldl@section{\mathversion{bold}#1}{#2}}
\makeatother

\makeatletter
\let\old@makecaption=\@makecaption
\def\@makecaption{\small\old@makecaption}
\makeatother

\begin{document}


\begin{flushright}\footnotesize
\texttt{NORDITA 2017-136} \\
\texttt{UUITP-52/17}
\end{flushright}

\renewcommand{\thefootnote}{\fnsymbol{footnote}}
\setcounter{footnote}{0}

\begin{center}
{\Large\textbf{\mathversion{bold} String corrections to circular Wilson loop\\ and anomalies}
\par}

\vspace{0.8cm}

\textrm{Alessandra Cagnazzo$^{1,2}$, Daniel Medina-Rincon$^{1,3}$ \\ and
Konstantin Zarembo$^{1,3}$\footnote{Also at ITEP, Moscow, Russia}}
\vspace{4mm}

\textit{${}^1$Nordita,  Stockholm University and KTH Royal Institute of Technology,
Roslagstullsbacken 23, SE-106 91 Stockholm, Sweden}\\
\textit{${}^2$Department of Physics, University of Oslo, P. O. Box 1048 Blindern, N-0316 Oslo,
Norway}\\
\textit{${}^3$Department of Physics and Astronomy, Uppsala University\\
SE-751 08 Uppsala, Sweden}\\
\vspace{0.2cm}
\texttt{zarembo@nordita.org}

\vspace{3mm}


\par\vspace{1cm}

\textbf{Abstract} \vspace{3mm}

\begin{minipage}{13cm}
We study string quantum corrections to the ratio of latitude and circular Wilson loops in $\mathcal{N}=4$ super-Yang-Mills theory at strong coupling. Conformal gauge for the corresponding minimal surface in $AdS_5\times S^5$ is singular and  we show that an IR anomaly associated with the divergence in the conformal factor removes previously reported discrepancy with the exact field-theory result. We also carefully check conformal anomaly cancellation and recalculate fluctuation determinants by directly evaluating phaseshifts for all the fluctuation modes.
\end{minipage}

\end{center}

\vspace{0.5cm}



\setcounter{page}{1}
\renewcommand{\thefootnote}{\arabic{footnote}}
\setcounter{footnote}{0}

\section{Introduction}

We will study the ratio of two Wilson loops in $\mathcal{N}=4$ super-Yang-Mills (SYM) theory that share a common contour in space-time, but differ in their coupling to scalars, following the proposal of \cite{Forini:2015bgo,Faraggi:2016ekd,Forini:2017whz}. Wilson loops are important observables in gauge theories and are unique probes of the AdS/CFT correspondence since they couple directly to the string worldsheet in the dual gravitational background \cite{Maldacena:1998im,Rey:1998ik,Drukker:1999zq}. Some Wilson loops in the SYM theory can be actually computed exactly, at any coupling strength, without making any approximations. Subsequent extrapolation to strong coupling establishes a direct link between conventional QFT calculations and holography. 

The simplest example of this type is the circular Wilson loop whose exact expectation value can be obtained by resumming diagrams of perturbation theory \cite{Erickson:2000af,Drukker:2000rr} or, at a more rigorous level, by localization of the path integral on $S^4$ \cite{Pestun:2007rz}. The strong-coupling extrapolation of the circle agrees precisely with the area law in $AdS_5\times S^5$, a result that can be generalized in many ways (see \cite{Zarembo:2016bbk} for a review). Quite surprisingly, even  the next order in the strong-coupling expansion has not been reproduced from string theory thus far, despite much effort \cite{Drukker:2000ep,Kruczenski:2008zk,Kristjansen:2012nz,Bergamin:2015vxa}, indicating that we do not understand in detail how strings in $AdS_5\times S^5$ and other holographic backgrounds should be quantized in the Wilson-loop sector.

The difficulty lies in the definition of the measure in the string path integral and in the delicate issues with reparameterization invariance on the string worldsheet. Taking the ratio of similar Wilson loops \cite{Forini:2015bgo,Faraggi:2016ekd} avoids these complications, because the measure factors simply cancel. For the ratio of the latitude and the circle, considered in \cite{Forini:2015bgo,Faraggi:2016ekd}, quantum string corrections can be computed exactly. Surprisingly, the result of the string calculation disagrees with the field-theory prediction \cite{Drukker:2006ga}. A different quantization prescription for string fluctuations \cite{Forini:2017whz} brings the result in agreement with field theory, but the method of \cite{Forini:2017whz} only applies to infinitesimally small deviations from the circle. The quantization prescriptions in \cite{Forini:2015bgo,Faraggi:2016ekd} and in \cite{Forini:2017whz} differ in the choice of the conformal frame on the string worldsheet, which {\it a priori} should not matter as long as the conformal anomaly cancels. 

We reconsider  string quantum corrections to the latitude Wilson loops, working in the same conformal frame as   \cite{Forini:2015bgo,Faraggi:2016ekd}. We  pay special attention to regularization issues and ensuing anomalies and will also carefully check that the conformal anomaly cancels, which is an important consistency condition in string theory.

\section{Circular Wilson loop and latitude}

\subsection{Latitude Wilson loops}

The Wilson loop expectation value in $\mathcal{N}=4$ SYM is defined as \cite{Maldacena:1998im}
\begin{equation}
 W(C;\mathbf{n})=\left\langle 
 \mathop{\mathrm{tr}}{\rm P}\exp\left[i\int_{C}^{}d\tau \,
 \left(\dot{x}^\mu A_\mu +i|\dot{x}|n^I\Phi _I\right)
 \right]
 \right\rangle,
\end{equation}
where $\Phi _I$ are the six scalar fields from the $\mathcal{N}=4$ supermultiplet and $\mathbf{n}$ is a unit six-dimensional vector that may change along the contour $C$. In string theory, the Wilson loop expectation value maps to the disc partition function with the boundary conditions determined by the contour $C=\left\{x^\mu (\tau )|\,\tau \in(0,2\pi )\right\}$ for the embedding coordinates in $AdS_5$, and by $\mathbf{n}(\tau )$ for $S^5$.

We concentrate on a particular family of  Wilson loops, for which $C$ is a unit circle and $\mathbf{n}$ is a latitude of $S^5$ \cite{Drukker:2006ga}: $\mathbf{n}=(\sin\theta _0\cos\tau ,\sin\theta _0\sin\tau ,\cos\theta _0,\mathbf{0})$. The expectation value of the latitude  is known exactly \cite{Drukker:2006ga}\footnote{The latitude
belongs to a more general class of supersymmetric Wilson loops which live on $S^2\in S^5$ and
reduce to the effective 2d Yang-Mills theory \cite{Drukker:2007dw} 
upon localization of the path integral \cite{Pestun:2009nn}.
The result quoted in the text is  large-$N$ exact.}:
\begin{equation}
 W(\theta _0)=\frac{2}{\sqrt{\lambda }\,\cos\theta _0}\,
 I_1\left(\sqrt{\lambda }\,\cos\theta _0\right),
\end{equation}
and interpolates between the simple circle at $\theta _0=0$ and a supersymmetric Wilson loop with trivial expectation value \cite{Zarembo:2002an} at $\theta _0=\pi /2$. Here $\lambda =g^2N$ is the 't~Hooft coupling of $\mathcal{N}=4$ SYM.
At strong coupling,
\begin{equation}\label{latitideWilson}
  W(\theta _0)=\sqrt{\frac{2}{\pi \cos^3\theta _0}}\,\lambda ^{-\frac{3}{4}}
  \,{\rm e}\,^{\sqrt{\lambda }\,\cos\theta _0}\left(1+\mathcal{O}\left(\lambda ^{-\frac{1}{2}}\right)\right).
\end{equation}
Notice that the strong coupling and BPS ($\theta _0\rightarrow \pi /2$) limits do not commute with one another.

In string theory, the exponent in the Wilson loop is determined by the area of the minimal surface with the given boundary conditions, while the prefactor is a contribution of the string fluctuations and of the measure in the string path integral. Following  \cite{Forini:2015bgo,Faraggi:2016ekd,Forini:2017whz}, we consider the ratio of  the circle to the latitude in which the complicated measure factor is expected to cancel\footnote{Another way to get rid of the measure factors is to consider infinitely stretched Wilson loops and concentrate on extensive quantities. In that case an agreement between field theory and quantum corrections in string theory was obtained for the quark-anti-quark potential in the $\mathcal{N}=4$ SYM \cite{Chu:2009qt,Forini:2010ek} and for the quark self-energy in the $\mathcal{N}=2^*$ theory \cite{Chen-Lin:2017pay}.}. The field-theory prediction for the Wilson loop ratio is
\begin{align}\label{localizationpredictiontheta}
 \Gamma \equiv \ln \frac{{{W\left( 0 \right)}  }}{{{W\left( {{\theta _0}} \right)}  }} = \sqrt \lambda  \left( { 1-\cos {\theta _0}} \right) +\frac{3}{2}\ln \cos {\theta _0} + \mathcal{O}\left( {{\lambda ^{ - 1/2}}} \right).
\end{align}
Our goal will be to reproduce this result from the explicit one-loop calculation in string theory.

\subsection{Classical solution}

In the standard Poincar\'e coordinates $\left\{x^\mu ,z\right\}$ of $AdS_5$ and in 
the angular coordinates $\theta ,\varphi $ of $S^2\subset S^5$, the minimal surface for the latitude is \cite{Drukker:2005cu}
\begin{eqnarray}\label{stringclassx}
 &&x^1=\frac{\cos\tau }{\cosh\sigma }\,,\qquad 
 x^2=\frac{\sin\tau }{\cosh\sigma }\,,\qquad
 z=\tanh\sigma \,,\qquad
\nonumber \\
 &&\cos\theta =\tanh(\sigma +\sigma _0),\qquad 
 \varphi =\tau ,
 \vphantom{x^1=\frac{\cos\tau }{\cosh\sigma }}
\end{eqnarray}
where $\sigma $ changes from $0$ to $\infty $ and $\theta _0$ is related to $\sigma _0$ as
\begin{equation}
 \tanh\sigma _0=\cos\theta _0.
\end{equation}
The induced worldsheet metric is given by
\begin{equation}\label{inducedmet}
d{s^2} = {\Omega ^2}\left( {d{\tau ^2} + d{\sigma ^2}} \right)
\end{equation}
with the scale factor 
$${\Omega ^2} = \frac{1}{{{{\sinh }^2}\sigma }} + \frac{1}{{{{\cosh }^2}\left( {\sigma  + {\sigma _0}} \right)}}\, .$$
Substituting the solution into the string action, and taking into account that the string tension is given by $\sqrt{\lambda }/2\pi $, in terms of the 't~Hooft coupling, one gets the correct exponent in (\ref{latitideWilson}). 

The field-theory prediction for the next, $\mathcal{O}(1)$  term in the  strong-coupling expansion of (\ref{localizationpredictiontheta}) is
\begin{equation}\label{1looplocalization}
 \Gamma _{\rm 1-loop}=\frac{3}{2}\,\ln\tanh\sigma _0.
\end{equation}
In string theory, this is expected to come from the one-loop quantum fluctuations of the string worldsheet \cite{Greensite:1999jw,Forste:1999qn,Kinar:1999xu,Drukker:2000ep}.

\subsection{One-loop string corrections}

The string oscillation modes around the classical solution (\ref{stringclassx}) are described by the following fluctuation operators  \cite{Forini:2015bgo,Faraggi:2016ekd,Forini:2017whz}\footnote{These can be obtained by specializing the general formalism of \cite{Drukker:2000ep,Forini:2015mca} to the classical solution (\ref{stringclassx}).}:
\begin{align}
{\widetilde{\mathcal{K}}_1} &=  - \partial _\tau ^2 - \partial _\sigma ^2 + \frac{2}{{{{\sinh }^2}\sigma }},\label{ScaledK1}\\
{\widetilde{\mathcal{K}}_2} &=  - \partial _\tau ^2 - \partial _\sigma ^2 - \frac{2}{{{{\cosh }^2}\left( {\sigma  + {\sigma _0}} \right)}},\label{ScaledK2}\\
\vphantom{\frac{2}{{{{\cosh }^2}\left( {\sigma  + {\sigma _0}} \right)}}}
{\widetilde{\mathcal{K}}_{3 \pm }} &=  - \partial _\tau ^2 - \partial _\sigma ^2 \pm 2i\left( {\tanh \left( {2\sigma  + {\sigma _0}} \right) - 1} \right){\partial _\tau }\nonumber\\ 
&\vphantom{\frac{2}{{{{\cosh }^2}\left( {\sigma  + {\sigma _0}} \right)}}}
\quad + \left( {\tanh \left( {2\sigma  + {\sigma _0}} \right) - 1} \right)\left( {1 + 3\tanh \left( {2\sigma  + {\sigma _0}} \right)} \right), \label{ScaledK3}\\
\label{rescaledFermion}
{{\widetilde{\mathcal{D}}}_\pm } & = i{\partial _\sigma } {\tau _1} - \left[{i{\partial _\tau } \mp \frac{1 }{2}\left( {1 - \tanh \left( {2\sigma  + {\sigma _0}} \right)} \right)} \right]{\tau _2}\nonumber\\
&\qquad  + \frac{1}{{{\Omega}\,{{\sinh }^2}\sigma }}\,{\tau _3} \mp\frac{1 }{{{\Omega}\,{{\cosh }^2}\left( {\sigma  + {\sigma _0}} \right)}}\,,
\end{align}
where $\tau _i$ are the standard Pauli matrices. The operator $\widetilde{\mathcal{K}}_1$ describes three string modes in $AdS_5$, the operator $\widetilde{\mathcal{K}}_2$ describes three modes on $S^5$,  $\widetilde{\mathcal{K}}_{3\pm}$ arise as a result of mixing between the two remaining modes -- one from the sphere, another from $AdS_5$. The Dirac operators $\widetilde{\mathcal{D}}_\pm $ originate from the kinetic terms for the eight fermions remaining after kappa-symmetry gauge-fixing in the Green-Schwarz action. 

The operators above are related to the ones that appear in the string action by a conformal transformation \cite{Forini:2015bgo}:
\begin{equation}\label{KGDir}
 \mathcal{K}=\frac{1}{\Omega ^2}\,\widetilde{\mathcal{K}},
\end{equation}
for bosons, and
\begin{equation}\label{f-trans}
 \mathcal{D}=\frac{1}{\Omega ^{\frac{3}{2}}}\widetilde{\mathcal{D}}\Omega ^{\frac{1}{2}},
\end{equation}
for fermions. The fluctuation modes of the string are naturally normalized with respect to the invariant measure of the induced metric (\ref{inducedmet}):
\begin{align}\label{innerproduct}
\left\langle {{{\phi _1}}}\!
 \mathrel{\left | {\vphantom {{{\phi _1}} {{\phi _2}}}}
 \right. \kern-\nulldelimiterspace}
 {{{\phi _2}}} \right\rangle  = \int d^2\sigma \, {\sqrt{h}\ {\phi^\dagger  _1}{\phi _2}}
 =\int_{}^{}d\tau \,d\sigma \,\Omega ^2{\phi ^\dagger _1}{\phi _2},
 \end{align}
and the fluctuation operators are Hermitian with respect to this scalar product, while the tilded operators are Hermitian with respect to the usual flat measure:
\begin{align}\label{flatinner}
\widetilde{\left\langle {{{\phi _1}}}\!
 \mathrel{\left | {\vphantom {{{\phi _1}} {{\phi _2}}}}
 \right. \kern-\nulldelimiterspace}
 {{{\phi _2}}} \right\rangle  }
 =\int_{}^{}d\tau \,d\sigma \,{\phi ^\dagger _1}{\phi _2}.
 \end{align}

The  one-loop partition function, that determines the Wilson loop expectation value, is given by the ratio of determinants of the physical, untilded operators \cite{Forini:2015bgo}:
\begin{equation}\label{Z-1-loop}
 Z(\sigma_{0})=\frac{{ {\det^{2} {{\mathcal{D}}_{+}}}\ {\det^{2} {{\mathcal{D}}_{-}}} }}{{{{\det }^{3/2}}{{\mathcal{K}}_1}\ {{\det }^{3/2}}{{\mathcal{K}}_2}\ {{\det }^{1/2}}{{\mathcal{K}}_{3 + }}\ {{\det }^{1/2}}{{\mathcal{K}}_{3 - }}}}.
\end{equation}
The Wilson loop is actually proportional to  $ Z(\sigma _0)$, but does not literally coincide with it. The string path integral contains some additional measure factors that are rather difficult to control. Fortunately, these factors do not depend on $\sigma _0$  and cancel in the ratio of the latitude to the circle. The one-loop free energy,  if normalized as in (\ref{localizationpredictiontheta}), 
 is given by the log-ratio of the partition functions:
\begin{equation}\label{string1-loop}
 \Gamma_{\rm 1-loop} =\ln\frac{Z(\infty )}{Z(\sigma_{0})}\,.
\end{equation}
This is the object we concentrate upon in the rest of the paper.

The tilded operators are technically easier to deal with, and in much of the previous work the conformal factors have been simply dropped.
Independence on the conformal frame is a basic principle of string theory. It is thus natural to assume that the untilded operators can be seamlessly replaced by the tilded ones. However, the log-ratio of determinants computed under this assumption (which we denote by $ \widetilde{\Gamma }_{\rm 1-loop}$) differs from the field-theory prediction (\ref{1looplocalization}) by an additional ``remainder" term \cite{Forini:2015bgo,Faraggi:2016ekd}:
\begin{equation}\label{wrong1-loop}
 \widetilde{\Gamma }_{\rm 1-loop}=
 \frac{3}{2}\,\ln  {\tanh {\sigma _0}}  -\frac{1}{2} \,\ln {\frac{{1 + \tanh {\sigma _0}}}{2}} \,.
\end{equation}

An obvious possible cause for the discrepancy, the one that first comes to mind, is the conformal anomaly. However it was argued in \cite{Faraggi:2016ekd} that the conformal anomaly is unlikely to account for the discrepancy. We refer to \cite{Faraggi:2016ekd,Drukker:2000ep} for technical details, and just remark that anomaly cancellation is very important in string theory. A non-zero contribution from the conformal anomaly would rather signal an internal inconsistency of the string calculation. 

\begin{figure}[t]
\begin{center}
 \subfigure[]{
   \includegraphics[width=6.5 cm] {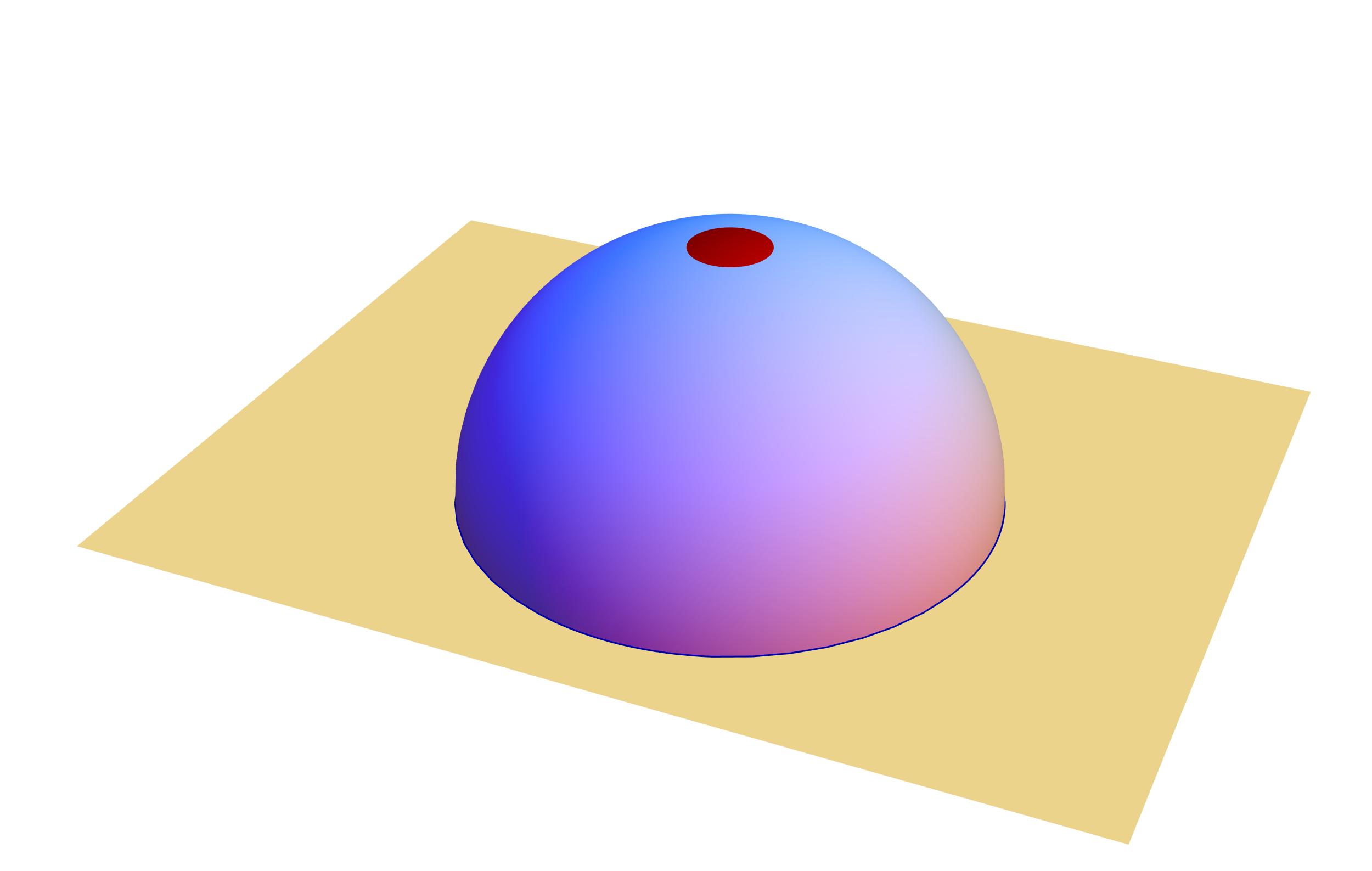}
   \label{fig2:subfig1}
 }
 \subfigure[]{
   \includegraphics[width=6.5 cm] {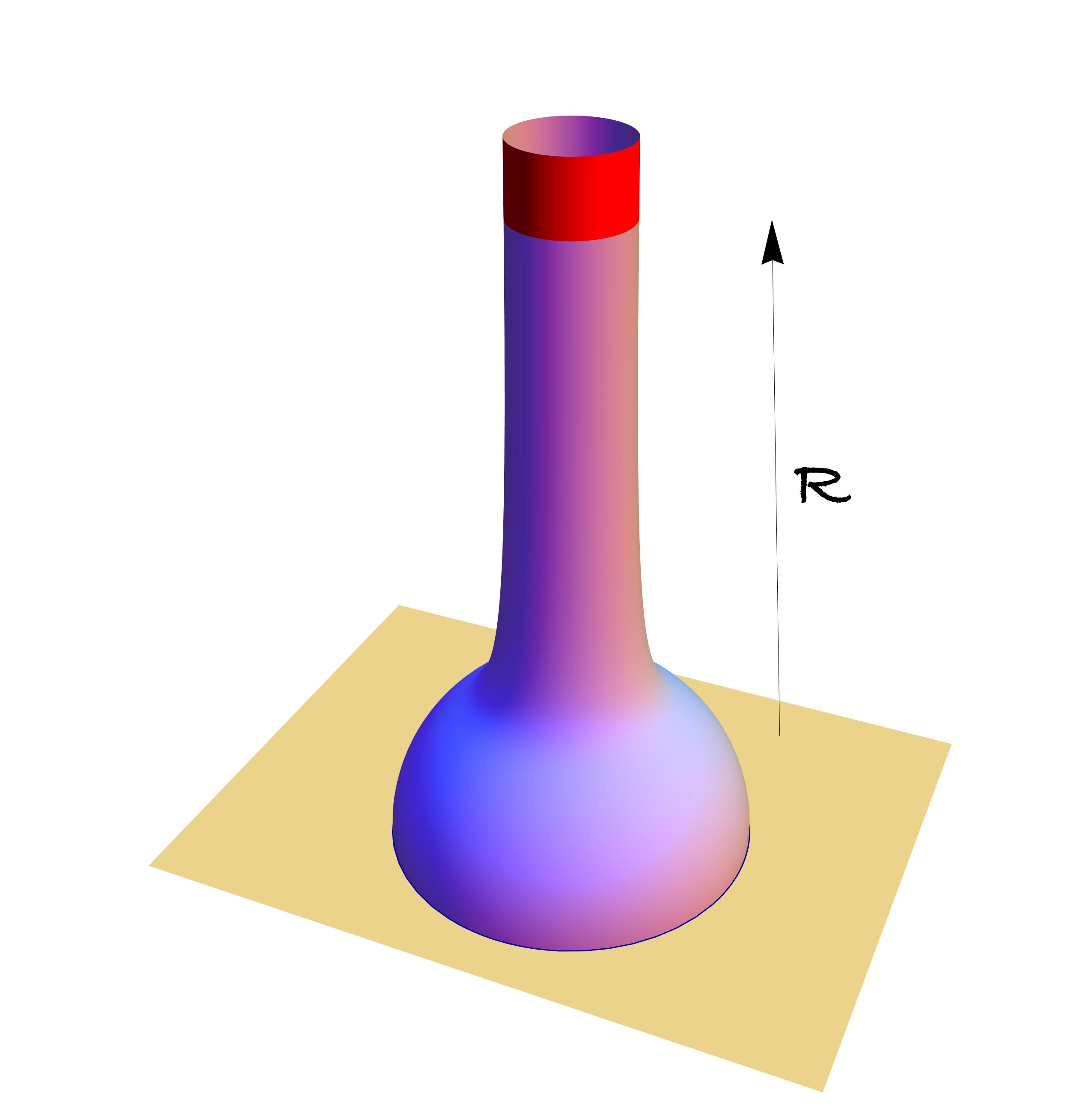}
   \label{fig2:subfig2}
 }
\caption{\label{AdSreg}\small Schematic representation of the induced metric on the minimal surface (a), and of the flat coordinates on the cylinder (b). The conformal transformation between the two is singular at the symmetry point of the minimal surface ($\sigma =\infty $). The region $\sigma >R$, removed by regularization, maps to  a small circle of area $s$ on the minimal surface in the target space.}
\end{center}
\end{figure}

A caveat here is that the conformal transformation from the metric of the disc (\ref{inducedmet}) to the flat metric of the semi-infinite cylinder changes the topology of the worldsheet and is actually singular at $\sigma =\infty $. The point $\sigma =\infty $ is regular in the induced metric (\ref{inducedmet}) but not in the flat metric, as illustrated in fig.~\ref{AdSreg}.  The spectral problem for a fluctuation operator on a cylinder differs from that on a disk in an essential way and requires an IR regularization. Even if the cutoff dependence eventually cancels out, regularization may leave a finite residue. We first give a simple but not very rigorous derivation of such an IR anomaly based on elementary thermodynamics, and then proceed with a more systematic analysis of the fluctuation determinants.

\subsection{Regularization and anomalies}

The change of the conformal frame of the form (\ref{KGDir}) corresponds to the following chain of transformations on the determinant of $\mathcal{K}$:
\begin{equation}\label{chain}
 \det\mathcal{K}=\left(\frac{\det\mathcal{K}}{\det\widetilde{\mathcal{K}}}\right)_{\rm anom.}\left(\frac{\det\widetilde{\mathcal{K}}}{\det\widetilde{\mathcal{K}}_\infty }\right)_{\rm cyl.}\det\widetilde{\mathcal{K}}_\infty ,
\end{equation}
where $\widetilde{\mathcal{K}}_\infty $ is the asymptotic operator obtained by taking $\sigma \rightarrow \infty $ in (\ref{ScaledK1})--(\ref{ScaledK3}), which is just the free Klein-Gordon operator:
\begin{equation}
 \widetilde{\mathcal{K}}_\infty =-\partial _\tau ^2-\partial _\sigma ^2,
\end{equation}
or the free Dirac operator, in case of fermions:
\begin{equation}
 \widetilde{\mathcal{D}}_\infty =i\tau _1\partial _\sigma -i\tau _2\partial _\tau .
\end{equation}

The first ratio in (\ref{chain}) is the conformal anomaly, the second one is well-defined on a cylinder, while separately $\det\widetilde{\mathcal{K}}$ and $\det\widetilde{\mathcal{K}}_\infty $ require an IR regularization. The IR cutoff is manifestly necessary for the Gel'fand-Yaglom method used in  \cite{Forini:2015bgo,Faraggi:2016ekd}, and is implicit in a more direct phaseshift calculation that will be carried out in sec.~\ref{detsandps}. 

The standard way to regularize the problem is to impose Dirichlet boundary conditions on the wavefunction of $\widetilde{\mathcal{K}}$ (or $\widetilde{\mathcal{K}}_\infty $) at some large but finite $\sigma =R$. This corresponds to removing a small segment of the minimal surface shown as a red circle in fig.~\ref{fig2:subfig1}. This is not such an innocent procedure, as can be seem by comparing determinants of the Laplacian on a disk and on a disk with a small hole \cite{Weisberger:1987kh}. Even though the IR cutoff cancels in the final answer,  intermediate steps do depend on $R$. At the same time,
the cutoff $R$ does not have any invariant meaning by itself.  To faithfully compare partition functions at different values of $\sigma _0$, we need a diffeomorphism-invariant regularization.

As an invariant regularization parameter we can take the area of the segment removed from the minimal surface when Dirichlet boundary conditions are imposed at $\sigma =R$:
\begin{equation}
 s=\int_{\sigma >R}^{}d^2\sigma \,\sqrt{h}
 =2\pi \int_{R}^{\infty }
 d\sigma\, \Omega ^2\simeq 
 4\pi \left(1+\,{\rm e}\,^{-2\sigma _0}\right)\,{\rm e}\,^{-2R}.
\end{equation}
The coordinate-dependent cutoff is related to the invariant one as
\begin{equation}\label{RR-inv}
 R=\frac{1}{2}\,\ln\frac{8\pi }{s\left(1+\tanh\sigma _0\right)}\equiv 
 R_{\rm inv}-\frac{1}{2}\,\ln\frac{1+\tanh\sigma _0}{2}\,.
\end{equation}
Because $R$ is a coordinate-dependent quantity with no invariant meaning, in comparing
the partition functions at different $\sigma _0$, it is $R_{\rm inv}$ rather than $R$ that should be kept fixed. Diffeomorphism-invariant regularization implies that $R$ has to be dependent on  $\sigma _0$.

The partition function depends on $R$ through the last factor in (\ref{chain}) and since $\widetilde{\mathcal{K}}_\infty $ and $\widetilde{\mathcal{D}}_\infty $ are just the free Klein-Gordon and Dirac operators, the asymptotic contribution to the partition function is given by the free energy of a gas of free particles in $1+1$ dimensions, given by
\begin{equation}
 F=-\left(2N_b+N_f\right)\,\frac{\pi }{12}\,T^2V.
\end{equation}
In our case $N_b=8=N_f$, $T=1/(2\pi )$ and $V=R$. We thus have
\begin{equation}
 \ln \widetilde{Z}_\infty =-\frac{F}{T}=R.
\end{equation}
The IR divergence cancels in the ratio (\ref{string1-loop}), but leaves a finite,  $\sigma _0$-dependent remnant due to (\ref{RR-inv}):
\begin{equation}
 \widetilde{\Gamma }_\infty =-R(\sigma _0)+R(\infty )=\frac{1}{2}\,\ln\frac{1+\tanh\sigma _0}{2}\,.
\end{equation}
Combined with (\ref{wrong1-loop}), this gives
\begin{equation}
 \widetilde{\Gamma }_{\rm 1-loop}+\widetilde{\Gamma }_\infty =
  \frac{3}{2}\,\ln  {\tanh {\sigma _0}} \,,
\end{equation}
which agrees with the localization prediction (\ref{1looplocalization}). 

This is the main result of the paper. To validate this result we need to check that the conformal anomaly cancels, which we do in the next section. Later we will also reanalyze the partition function on the cylinder and will derive the above result by a direct spectral analysis of the fluctuation operators.
 
\section{Conformal anomaly cancellation}

The anomaly contribution to the free energy is
\begin{equation}\label{Gamma-anom}
 \Gamma _{\rm anom}=\frac{1}{2}\sum_{a}^{}\left(-1\right)^{F_a}
 \ln\frac{\det\mathcal{K}_a}{\det\widetilde{\mathcal{K}}_a}\, ,
\end{equation}
where the summation runs over all operators in (\ref{Z-1-loop}) with the appropriate multiplicities. The operators in the numerator and denominator differ by the conformal factor, and if one were allowed to factorize the determinants, the sum would trivially vanish. The story is more complicated because the determinants need  intermediate UV regularization, which leaves a finite remnant, the anomaly. 

The anomaly, being a local effect of  UV divergences, can be computed by the standard DeWitt-Seeley expansion. For completeness we give a brief derivation of the conformal anomaly adapted to our case in the appendix~\ref{conformalan-app}. The results (\ref{b-anomaly}) and (\ref{f-anomaly}) directly apply to the operators (\ref{ScaledK1})--(\ref{f-trans}), upon bringing them to the standard Klein-Gordon/Dirac form (\ref{K(alpha)}), (\ref{D(alpha)}), with the identifications
\begin{eqnarray}
\phi&=&-\ln\Omega 
\vphantom{\frac{2}{\sinh^2\sigma }}\, ,
\nonumber \\
 E_1&=&\frac{2}{\sinh^2\sigma },
\nonumber \\
E_2&=&-\frac{2}{\cosh^2(\sigma +\sigma _0)},
\nonumber \\
E_{3\pm}&=&-\frac{2}{\cosh^2(2\sigma +\sigma _0)},
\nonumber \\
 a_\pm&=&\frac{1}{\Omega \sinh^2\sigma },
\nonumber \\
 v_\pm&=&\mp\frac{1}{\Omega \cosh^2(\sigma +\sigma _0)}\,.
\end{eqnarray}

First, we notice that the boundary terms in the anomaly trivially cancel between bosons and fermions, just by matching the number of degrees of freedom. To see that the bulk anomaly also cancels it is convenient to bring the scale factor of the metric to the following form:
\begin{equation}
 \Omega ^2=\frac{\cosh\sigma _0\cosh(2\sigma +\sigma _0)}{\sinh^2\sigma \cosh^2(\sigma +\sigma _0)}\,,
\end{equation}
from which it immediately follows that
$$
 \partial _\sigma ^2\phi =-\partial _\sigma ^2\ln\Omega =
 \frac{1}{\cosh^2(\sigma +\sigma _0)} -\frac{1}{\sinh^2\sigma }
 -\frac{2}{\cosh^2(2\sigma +\sigma _0)}\,.
$$
This enters the anomaly with a prefactor
$$
\left(\frac{1}{6}\times 8+\frac{1}{12}\times 8\right)\phi  =2\phi.  
$$
On the other hand,
\begin{eqnarray}
 &&3E_1+3E_2+E_{3+}+E_{3-}+4\left(v_+^2-a_+^2\right)+4\left(v_-^2-a_-^2\right)
\nonumber \\
&&=\frac{2}{\cosh^2(\sigma +\sigma _0)}-\frac{2}{\sinh^2\sigma }-\frac{4}{\cosh^2(2\sigma +\sigma _0)}\,,
\nonumber
\end{eqnarray}
and the two terms in the anomaly completely compensate one another.

The anomaly thus cancels in the partition function (\ref{Z-1-loop}) at each value of $\sigma _0$, not only in the ratio, as actually expected.

\section{Determinants and phaseshifts}\label{detsandps}

The fluctuation determinants were evaluated in \cite{Forini:2015bgo,Faraggi:2016ekd} with the help of the Gel'fand-Yaglom method. Here we recalculate them by a more direct approach, evaluating phaseshifts for each operator and then integrating over the phase space of string fluctuations. First we set up the general scheme for the phaseshift computation and then apply it to each operator in turn.

\subsection{Preliminaries}\label{SecPreliminaries}

The operators at hand have the general form (we start with bosons, for fermions the same scheme works with minor modifications):
\begin{equation}\label{KOpform1}
 \widetilde{\mathcal{K}}=-\partial _\sigma ^2+V(\partial _\tau ,\sigma ).
\end{equation}
The asymptotics at infinity are that of the free d'Alambert operator:
\begin{equation}
 V(\partial _\tau ,\infty )=-\partial _\tau ^2.
\end{equation}
The Fourier expansion in $\tau $ replaces $\partial _\tau $ by $-i\omega $, with integer frequency, or half-integer depending on whether the boundary conditions are periodic or anti-periodic in the $\tau $ direction. The spectrum thus decomposes into a sequence of one-dimensional problems for each Fourier mode:
\begin{equation}\label{basicSch}
 \left(-\partial _\sigma ^2+V(-i\omega,\sigma  )\right)\Psi =\Lambda \Psi .
\end{equation}

The boundary condition at $\sigma =0$ is
\begin{equation}\label{goodbound}
 \Psi (0)=0.
\end{equation}
After the boundary condition is imposed the wavefunction is fixed up to normalization. Since the potential vanishes at infinity, the wavefunction asymptotically has an oscillating behaviour:
\begin{equation}\label{asymptwave}
 \Psi (\sigma )
 \stackrel{\sigma \rightarrow \infty }{\simeq }
 {\rm C}\sin(p\sigma +\delta ).
\end{equation}
The eigenvalue can be read off the asymptotic form of the Schr\"odinger equation (\ref{basicSch}):
\begin{equation}
 \Lambda =\omega ^2+p^2.
\end{equation}

To define the determinant of an operator with a continuous spectrum we need to introduce an IR cutoff by imposing another boundary condition at $\sigma =R$: 
\begin{equation}\label{extra-bc}
 \Psi (R)=0. 
\end{equation}
The spectrum then becomes discrete due to momentum quantization condition:
\begin{equation}\label{quantizationcond}
 p_nR+\delta (\omega ,p_n)\simeq \pi n,
\end{equation}
which follows from the asymptotic form of the wavefunction (\ref{asymptwave}) and is thus valid as long as $R$ is much larger than the range of the potential in 
the Schr\"odinger equation. The density of states $\rho =\partial n/\partial p$ in the limit of $R\rightarrow \infty $ hence takes the form
\begin{equation}
 \rho (p)=\frac{1}{\pi }\left(R+\frac{\partial \delta (\omega ,p)}{\partial p}\right).
\end{equation}
The often omitted extensive piece proportional to $R$ has to be kept here. The infrared divergence cancels in the ratio of the two partition functions at different $\sigma _0$.  But we have seen that the cutoff $R$ depends on $\sigma _0$ if regularization is to preserve general covariance, and therefore the extensive part of the partition function has to be kept throughout the calculation.

The determinant of $\widetilde{\mathcal{K}}$ is obtained by multiplying all the eigenvalues:
\begin{eqnarray}
 \ln \det \widetilde{\mathcal{K}}&=&\sum_{\omega}^{}\int_{0}^{\infty }\frac{dp}{\pi }\,\left(R+
 \frac{\partial \delta (\omega ,p)}{\partial p}\right)\ln(\omega ^2+p^2)
\label{eq39}
\nonumber \\
 &=&-\sum_{\omega}^{}\int_{0}^{\infty }\frac{dp}{\pi }\,\,
 \frac{2p}{\omega ^2+p^2}\,\left(\delta (\omega ,p)+Rp\right),
\end{eqnarray}
with $\omega\in\mathbb{Z}$ or $\omega\in\mathbb{Z}+1/2$ for periodic/anti-periodic boundary conditions.

Our strategy will be to directly evaluate phaseshifts for all the operators, sum over frequencies and integrate over spacial momenta. Before proceeding with explicit calculations we sum over the Matsubara frequency using standard tools of Statistical Mechanics \cite{AGD}, and make a few technical remarks that streamline  calculation of the phaseshifts.

\subsection{Summation over frequencies}

\begin{figure}[t]
\begin{center}
 \centerline{\includegraphics[width=6cm]{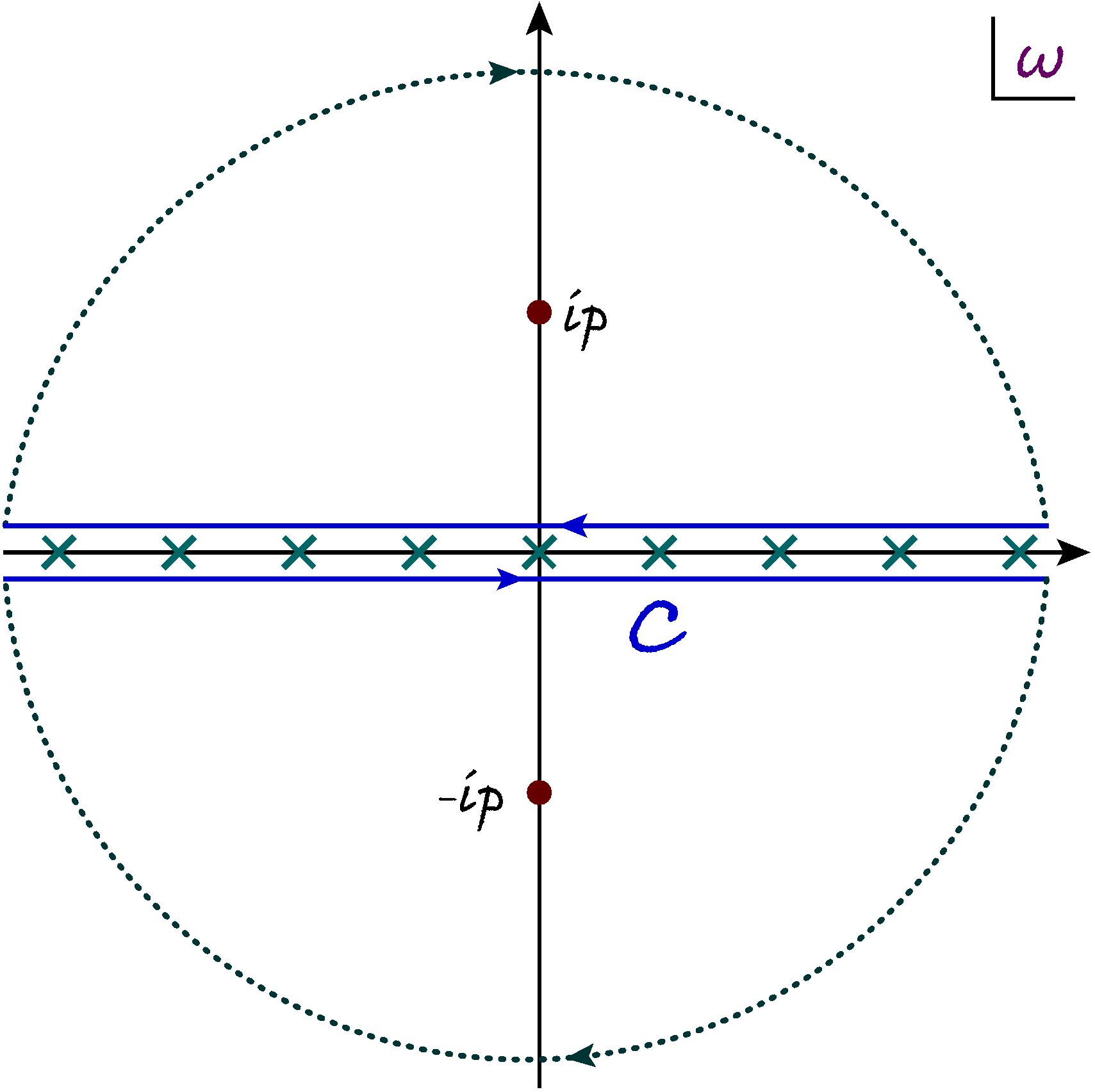}}
\caption{\label{contours:fig}\small Contours of integration in the complex frequency plane.}
\end{center}
\end{figure}

The standard trick is to replace summation  by integration along the contour  shown in fig.~\ref{contours:fig}:
\begin{equation}
\ln \det \widetilde{\mathcal{K}} =  - \int_C^{} {\frac{{d\omega }}{{2\pi i}}} \,\,\cot {\pi \omega } \,\int_0^\infty  {dp} \,\,\frac{{2p}}{{{\omega ^2} + {p^2}  }}\,
\left(\delta (\omega ,p)+Rp\right).
\end{equation}
The poles of the cotangent recover the sum over the Matsubara frequencies.

Assuming that the phaseshift does not grow too fast at large frequencies (in the simplest cases the phaseshift does not depend on the frequency at all), the contour of integration can be closed in the upper and lower half-planes, as shown in fig.~\ref{contours:fig}, after which the integral over $\omega $ picks up residues at $\omega =\pm ip$. 
 Denoting
\begin{equation}
 \delta _\pm(p)\equiv \delta (\pm ip,p), 
\end{equation}
we get for the determinant:
\begin{equation}
\ln \det \widetilde{\mathcal{K}} =  - \int_0^\infty  {dp} \,\coth \pi p \,\left( {{\delta _ + }(p) + {\delta _ - }(p)}+2Rp \right).\label{lnKB}
\end{equation}
This equation expresses the determinant entirely through the on-shell data. It suffices to solve the Schr\"odinger equation (\ref{basicSch}) at $\omega =\pm ip$ and $\Lambda =0$. 

If the boundary conditions are anti-periodic and the Matsubara frequencies $\omega$ are half-integer, the summation formulas differ by substitutions $\cot\rightarrow -\tan$, $\coth\rightarrow \tanh$:
\begin{equation}
\ln \det {\widetilde{\mathcal{K}}_F} =  - \int_0^\infty  {dp} \,\,\tanh \pi p\,\left( {{\delta _ + }(p) + {\delta _ - }(p)} +2Rp\right).\label{lnKF}
\end{equation}

Normally, the particle and anti-particle phaseshifts $\delta _+$ and $\delta _-$ are equal, but some operators that we encounter have a {\it spectral asymmetry} resulting in different density of states for particles and anti-particles. 

\subsection{Phaseshifts and Jost functions}

Instead of solving the Schr\"odinger equation with the correct boundary conditions (\ref{goodbound}) it is sometimes easier to find the Jost functions, which are the solutions that asymptote to unit-normalized plane waves at infinity:
\begin{equation}
 Y_p(\sigma )\stackrel{\sigma \rightarrow \infty }{\simeq }\,{\rm e}\,^{ip\sigma },
 \qquad 
 \bar{Y}_p(\sigma )\stackrel{\sigma \rightarrow \infty }{\simeq }\,{\rm e}\,^{-ip\sigma }.
\end{equation}
For a self-adjoint Schr\"odinger operator with a real potential the Jost functions are complex conjugate to one another: $\bar{Y}_p=Y_p^*$, but if the potential is complex, which is the case for the operators $\widetilde{\mathcal{K}}_{3\pm}$ for example, then the two Jost functions are not related in any simple way.

The Jost functions form a complete set of solutions to the Schr\"odinger equation. The solution that satisfies the correct boundary conditions is a linear combination of the two Jost functions:
\begin{equation}
 \Psi _p(\sigma )=\bar{Y}_p(0)Y_p(\sigma )-Y_p(0)\bar{Y}_p(\sigma ).
\end{equation}
This linear combination indeed vanishes at $\sigma =0$. Comparing its behaviour at infinity with (\ref{asymptwave}), we find:
\begin{equation}
 \bar{Y}_p(0)\,{\rm e}\,^{ip\sigma }-Y_p(0)\,{\rm e}\,^{-ip\sigma }
 =
 \frac{{\rm C}}{2i}\,\,{\rm e}\,^{ip\sigma +\delta }- \frac{{\rm C}}{2i}\,\,{\rm e}\,^{-ip\sigma -\delta },
\end{equation}
which expresses the phaseshift through the Jost  data:
\begin{equation}
 \frac{\bar{Y}_p(0)}{Y_p(0)}=\,{\rm e}\,^{2i\delta }\,.
\end{equation}

In the self-adjoint case the Jost functions are complex conjugate and their ratio is a pure phase. The phaseshift is real in this case. If the Schr\"odinger operator is not self-adjoint, the phaseshift may have an imaginary part. In general,
\begin{equation}\label{Jost->delta}
 \delta (p)=\frac{i}{2}\,\ln\frac{{Y}_p(0)}{\bar{Y}_p(0)}\,.
\end{equation}
We will use this formula to evaluate the phaseshifts of the fluctuation operators for the latitude by explicitly calculating the Jost functions in each case. The same scheme can be applied to fermions with minor modifications related to their two-component nature.

\subsection{The phaseshift computation}

\subsubsection{Operator $\widetilde{\mathcal{K}}_{1}$}
The differential equation for this operator is given by
\begin{align}
\left(- \partial _\sigma ^2 + \frac{2}{{{{\sinh }^2}\sigma }}\right){\chi_1} = {p^2}{\chi_1}.
\end{align}
The solutions to this equation are given by the Jost functions
\begin{align}
{Y_p}\left( \sigma  \right) = {\,{\rm e}\,^{ip\sigma }}\frac{{ip - \coth \sigma }}{{ip - 1}}, && {{\bar Y}_p}\left( \sigma  \right) =   {\,{\rm e}\,^{ - ip\sigma }}\frac{{ip + \coth \sigma }}{{ip + 1}},
\end{align}
satisfying
 ${Y_p} = {\left( {{{\bar Y}_p}} \right)^*}$ and ${{\bar Y}_p} = {\left( {{Y_p}} \right)^*}$.
Using equation \eqref{Jost->delta}, we obtain
\begin{align}\label{delta1}
 {\delta _1} = \frac{i}{2}\ln  {\frac{{ ip+1}}{{ ip-1}}} =\frac{\pi }{2}-\arctan p.
\end{align}

\subsubsection{Operator $\widetilde{\mathcal{K}}_{2}$}
The corresponding differential equation is given by
\begin{align}
 \left(- \partial _\sigma ^2 - \frac{2}{{{{\cosh }^2}\left( {\sigma  + {\sigma _0}} \right)}}\right){\chi_2} = {p^2}{\chi_2}.
\end{align}
In this case, the Jost functions are
\begin{align}
{Y_p} = {\,{\rm e}\,^{ip\sigma }}\frac{{ip - \tanh \left( {\sigma  + {\sigma _0}} \right)}}{{ip - 1}}, && {{\bar Y}_p} = {\,{\rm e}\,^{ - ip\sigma }}\frac{{ip + \tanh \left( {\sigma  + {\sigma _0}} \right)}}{{ip + 1}},
\end{align}
satisfying ${Y_p} = {\left( {{{\bar Y}_p}} \right)^*}$ and ${{\bar Y}_p} = {\left( {{Y_p}} \right)^*}$. From equation \eqref{Jost->delta}, we have
\begin{align}\label{delta2}
{\delta _2} = \frac{i}{2}\ln \left( {\frac{{1 + ip}}{{1 - ip}}\frac{{\tanh {\sigma _0} - ip}}{{\tanh {\sigma _0} + ip}}} \right) =  - \arctan  p + \arctan{\frac{p}{{\tanh {\sigma _0}}}} \,.
\end{align}

\subsubsection{Operator $\widetilde{\mathcal{K}}_{3\pm}$}

The Schr\"odinger problem for $\widetilde{\mathcal{K}}_{3+}$ is
\begin{equation}\label{RosenMorse}
 \left[
-\partial _\sigma ^2+\left(3\tanh(2\sigma +\sigma _0)+1\pm 2ip\right) 
\left(\tanh(2\sigma +\sigma _0)-1\right)
 \right]\psi _p=p^2\psi _p,
\end{equation}
where we have set  $\partial _\tau \rightarrow -i\omega =\pm p$.  The $\pm $ sign refers to particle/anti-particle modes.

The potential in the Schr\"odinger equation is of the solvable Rosen-Morse type. The  solution\footnote{The Rosen-Morse potential is solvable in hypergeometric functions for any values of the frequency $\omega $, not necessarily on-shell \cite{rosen1932vibrations}. Using this general solution an analytic expression for the off-shell phaseshift $\delta (\omega ,p)$ can be found for any $\omega $ and $p$. We do not display this more general function here, because we only need the on-shell phaseshifts $\delta _\pm(p)$ to compute the determinant.} can be found by the substitution  \cite{rosen1932vibrations}
\begin{equation}
 x=\frac{1-\tanh(2\sigma +\sigma _0)}{2}
\end{equation}
accompanied by the following ansatz for the wavefunction:
\begin{equation}
 \psi _p(\sigma )=\,{\rm e}\,^{\mp ip\sigma +\sigma }\cosh^{-\frac{1}{2}}(2\sigma +\sigma _0)\chi (x),
\end{equation}
which leads to
\begin{equation}
 \left[\frac{d}{dx}x\left(1-x\right)\frac{d}{dx}-\left(x\mp\frac{ip}{2}\right)\frac{d}{dx}\right]\chi =0,
\end{equation}
or
\begin{equation}
 \psi _p(\sigma )=\,{\rm e}\,^{\pm ip\sigma -\sigma }\cosh^{\frac{1}{2}}(2\sigma +\sigma _0)\tilde{\chi }(x),
\end{equation}
which leads to
\begin{equation}
 \left[\frac{d}{dx}x\left(1-x\right)\frac{d}{dx}+\left(x\mp\frac{ip}{2}\right)\frac{d}{dx}+1\right]\tilde{\chi } =0.
\end{equation}
These equations have simple solutions:
\begin{equation}
 \chi (x)=1,\qquad 
 \tilde{\chi }(x)=x-1\pm\frac{ip}{2}\,.
\end{equation}
From here we find the Jost functions:
\begin{eqnarray}\label{Jost3+}
 &&Y_{p}^+(\sigma )=\,{\rm e}\,^{ip\sigma -\sigma -\frac{\sigma _0}{2}}
 \sqrt{2\cosh(2\sigma +\sigma _0)}\,\,\frac{1-ip+\tanh(2\sigma +\sigma _0)}{2-ip},
\nonumber \\
 &&\bar{Y}_{p}^+(\sigma )=\frac{
 \,{\rm e}\,^{-ip\sigma +\sigma +\frac{\sigma _0}{2}}}
 {\sqrt{2\cosh(2\sigma +\sigma _0)}},
\nonumber \\
\vphantom{\bar{Y}_{p}^+(\sigma )=\frac{
 \,{\rm e}\,^{-ip\sigma +\sigma +\frac{\sigma _0}{2}}}
 {\sqrt{2\cosh(2\sigma +\sigma _0)}}}
&& Y_{p}^-=\left(\bar{Y}_{p}^+\right)^*,\qquad \bar{Y}_{p}^-=\left(Y_{p}^+\right)^*.
\end{eqnarray}

The Jost functions $Y_{p}$ and $\bar{Y}_{p}$ are not complex conjugate to one another, because the potential  is complex, and consequently the phaseshifts will have an imaginary part. There is also a spectral asymmetry between particle and anti-particle phaseshifts, but it is easy to quantify it because particle and anti-particle Jost functions are related by complex conjugation: $\delta _-(p)^*=\delta _+(p)$. Taking into account the last equation in (\ref{Jost3+}), we get from (\ref{Jost->delta}):
\begin{equation}
 \delta _{3+}(p)+\delta _{3-}(p)=\frac{i}{2}\,\ln\frac{Y^+_p(0)\bar{Y}^+_p(0)^*}{Y^+_p(0)^*\bar{Y}_p^+(0)}=\arctan\frac{p}{1+\tanh\sigma _0}-\arctan\frac{p}{2}\,.\label{delta3}
\end{equation}
The answer for $\widetilde{\mathcal{K}}_{3-}$ is the same up to exchanging $\delta _{3+}$ with $\delta _{3-}$.

\subsubsection{Operator $\widetilde{\mathcal{D}}_{\pm}$}\label{spectralproblemFermi}

It is convenient to consider, instead of $\widetilde{\mathcal{D}}_{\alpha}$, the eigenvalue problem for $i\tau_{2}\widetilde{\mathcal{D}}_{\alpha}$. The index $\alpha $ that takes values $\pm$ is introduced here in order to distinguish the operator label from the particle/anti-particle index.  The spectral problem for the resulting Dirac operator takes the form:
\small{\begin{equation}\label{eigenvalueproblemfermio}
\left(i{\tau _3}{\partial _\sigma } + \frac{i\alpha }{2}\left[ {1 - \tanh \left( {2\sigma  + {\sigma _0}} \right)} \right]\mathbbm{1} - \frac{1}{{\Omega {{\sinh }^2}\sigma }}{\tau _1} - i\frac{\alpha }{{\Omega {{\cosh }^2}\left( {\sigma  + {\sigma _0}} \right)}}{\tau _2}\right){\chi _\alpha } = \mp p{\chi _\alpha } ,
\end{equation}}
where the $\mp$ sign comes from the two choices of closing the integration contour described in figure \ref{contours:fig}. 

The most general solution of the Dirac equation is a superposition of the Jost functions
\small{\begin{align}
Y_{p,\alpha }^ \pm &= {e^{ \pm \alpha ip\sigma }}\left( {{\delta _{\alpha , + }}\left( {\begin{array}{*{20}{c}}
  {c_{\text{I}}} \\ 
  {c_{\text{II}}} 
\end{array}} \right) + {\delta _{\alpha , - }}\left( {\begin{array}{*{20}{c}}
  {c_{\text{II}}} \\ 
  {c_{\text{I}}} 
\end{array}} \right)} \right),\\
\bar Y_{p,\alpha }^ \pm  &= {e^{ \mp \alpha ip\sigma }}\left( {{\delta _{\alpha , + }}\left( {\begin{array}{*{20}{c}}
  {\bar{c}_{\text{I}}} \\ 
  {\bar{c}_{\text{II}}} 
\end{array}} \right) + {\delta _{\alpha , - }}\left( {\begin{array}{*{20}{c}}
  {\bar{c}_{\text{II}}} \\ 
  {\bar{c}_{\text{I}}} 
\end{array}} \right)} \right),
\end{align}}
where
\small{
\begin{align}
c_{\text{I}} &= \frac{1}{{\left( { \pm \alpha ip - \frac{3}{2}} \right)\left( { \pm \alpha ip - \frac{1}{2}} \right)}}\frac{{{2^{ - 7/4}}{e^{\sigma /2}}{e^{ - 5{\sigma _0}/4}}{\Omega ^{1/2}}}}{{{{\cosh }^{1/4}}{\sigma _0}\sqrt {\sinh \sigma \cosh \left( {\sigma  + {\sigma _0}} \right)} }}\nonumber\\
&\left[ {{e^{ - 3\sigma }}\left( {{p^2} + \frac{1}{4}} \right) + \left( { \pm \alpha ip - \frac{3}{2}} \right)\left( {{e^{ - \sigma }}\left( { \pm \alpha ip + \frac{1}{2}} \right) + 2{e^{2{\sigma _0}}}\sinh \sigma \left( { \pm \alpha ip - \frac{{\coth \sigma }}{2}} \right)} \right)} \right],\nonumber\\
c_{\text{II}} &= \frac{{i\alpha }}{{\left( { \pm \alpha ip - \frac{3}{2}} \right)\left( { \pm \alpha ip - \frac{1}{2}} \right)}}\frac{{{2^{ - 7/4}}{e^{ - \sigma /2}}{e^{ - {\sigma _0}/4}}\Omega }}{{\cosh {\sigma _0}{{\cosh }^{3/4}}\left( {2\sigma  + {\sigma _0}} \right)}}\nonumber\\
&\left[ {\left( { \pm \alpha ip - \frac{1}{2}} \right)\left( {2 + \cosh \left( {2\left( {\sigma  + {\sigma _0}} \right)} \right) - \cosh \left( {2\sigma } \right)} \right) - \sinh \left( {2\left( {\sigma  + {\sigma _0}} \right)} \right) + \sinh \left( {2\sigma } \right)} \right],\nonumber\\
\bar{c}_{\text{I}} &= \frac{{i\alpha }}{{\left( { \pm \alpha ip + \frac{1}{2}} \right)}}\frac{{{e^{\sigma /2}}{e^{{\sigma _0}/4}}\Omega }}{{{2^{5/4}}{{\cosh }^{1/4}}\left( {2\sigma  + {\sigma _0}} \right)}},\nonumber\\
\bar{c}_{\text{II}} &= \frac{1}{{\left( { \pm \alpha ip + \frac{1}{2}} \right)}}\frac{{{e^{\sigma /2}}{e^{{\sigma _0}/4}}}}{{{2^{1/4}}{{\cosh }^{1/4}}\left( {2\sigma  + {\sigma _0}} \right)}}\left( { \pm \alpha ip + \frac{1}{2}\left( {\frac{{\cosh \left( {2\sigma  + {\sigma _0}} \right)}}{{\sinh \sigma \cosh \left( {\sigma  + {\sigma _0}} \right)}} - 1} \right)} \right).\nonumber
\end{align}}
Asymptotically, the two solutions behave as
\begin{align*}
\mathop {\lim }\limits_{\sigma  \to \infty } Y_{p,\alpha }^ \pm  &= {e^{ \pm \alpha ip\sigma }}\left( {{\delta _{\alpha , + }}\left( {\begin{array}{*{20}{c}}
  1 \\ 
  0 
\end{array}} \right) + {\delta _{\alpha , - }}\left( {\begin{array}{*{20}{c}}
  0 \\ 
  1 
\end{array}} \right)} \right),\\
\mathop {\lim }\limits_{\sigma  \to \infty } \bar Y_{p,\alpha }^ \pm  &= {e^{ \mp \alpha ip\sigma }}\left( {{\delta _{\alpha , + }}\left( {\begin{array}{*{20}{c}}
  0 \\ 
  1 
\end{array}} \right) + {\delta _{\alpha , - }}\left( {\begin{array}{*{20}{c}}
  1 \\ 
  0 
\end{array}} \right)} \right).
\end{align*}

Close to $\sigma=0$, the Jost functions behave as
\begin{align*}
\mathop {\lim }\limits_{\sigma  \to 0} Y_{p,\alpha }^ \pm  &= \frac{v_{\alpha}^{\pm}}{\sigma }\left( {{\delta _{\alpha , + }}\left( {\begin{array}{*{20}{c}}
  i \\ 
  1 
\end{array}} \right) + {\delta _{\alpha , - }}\left( {\begin{array}{*{20}{c}}
  1 \\ 
  { - i} 
\end{array}} \right)} \right) + \mathcal{O}\left( \sigma  \right),\\
\mathop {\lim }\limits_{\sigma  \to 0} \bar Y_{p,\alpha }^ \pm  &= \frac{{\bar v}_{\alpha}^{\pm}}{\sigma }\left( {{\delta _{\alpha , + }}\left( {\begin{array}{*{20}{c}}
  i \\ 
  1 
\end{array}} \right) + {\delta _{\alpha , - }}\left( {\begin{array}{*{20}{c}}
  1 \\ 
  { - i} 
\end{array}} \right)} \right) + \mathcal{O}\left( \sigma  \right),
\end{align*}
where
\begin{align*}
v_{\alpha}^{\pm} = i\alpha \frac{{{{\cosh }^{1/4}}{\sigma _0}}}{{{2^{3/4}}{e^{{\sigma _0}/4}}}}\frac{{ { \pm \alpha ip -  {\frac{1}{2} - \tanh {\sigma _0}} } }}{{\left( { \pm \alpha ip - \frac{3}{2}} \right)\left( { \pm \alpha ip - \frac{1}{2}} \right)}},&&
\bar{v}_{\alpha}^{\pm} = \frac{{{e^{{\sigma _0}/4}}}}{{{2^{5/4}}{{\cosh }^{1/4}}{\sigma _0}}}\frac{1}{{ { \pm \alpha ip + \frac{1}{2}} }}.
\end{align*}
From the above, it is easy to see that the superposition $$
\chi _\alpha ^\pm =\bar v_\alpha ^ \pm Y_{p,\alpha }^ \pm  - v_\alpha ^ \pm \bar Y_{p,\alpha }^ \pm$$ 
vanishes for $\sigma\rightarrow0$, and therefore
satisfies the right boundary conditions at $\sigma =0$. At $\sigma \rightarrow \infty $, the correct solution behaves as
\begin{equation}
 \chi _\alpha ^\pm\simeq 
 \delta _{\alpha ,+}\begin{pmatrix}
  \bar{v}_+ ^\pm\,{\rm e}\,^{\pm i p\sigma }  \\ 
  -v_+ ^\pm\,{\rm e}\,^{\mp ip\sigma } \\ 
 \end{pmatrix}
 +\delta _{\alpha ,-}\begin{pmatrix}
  -{v}_- ^\pm\,{\rm e}\,^{\pm i p\sigma }  \\ 
  \bar{v}_- ^\pm\,{\rm e}\,^{\mp i p\sigma } \\ 
 \end{pmatrix}.
\end{equation}

To define the fermionic phaseshift we need to understand what replaces the auxiliary boundary condition (\ref{extra-bc}) for fermions. Since the Dirac equation is of the first order it is impossible to set both spinor components of the wavefunction to zero. Only a chiral projection of the wavefunction can vanish. Choosing the chirality condition  (any other choice leads to equivalent results) as
\begin{equation}
 \tau _2\chi (R)=\chi (R),
\end{equation}
we get the momentum quantization condition in the form (\ref{quantizationcond}) with
\begin{align}\label{deltaalphaF}
\delta _\alpha ^ \pm  = \pm \frac{\alpha }{{ 2}}\text{Arg}\left( {\frac{{\bar v_\alpha ^ \pm }}{{v_\alpha ^ \pm }}} \right) =\frac{\pi }{2}+\frac{1}{2}\arctan {\frac{p}{{\frac{1}{2} + \tanh {\sigma _0}}}}   - \arctan {2p}  - \frac{1}{2}\arctan {\frac{{2p}}{3}} \,. 
\end{align}

\subsection{Collecting the pieces together}

Expressing the determinants in (\ref{Z-1-loop}) through the on-shell phaseshifts with the help of (\ref{lnKB}) and (\ref{lnKF}), and collecting all the pieces together we get for the log of the partition function:
\begin{eqnarray}
 \ln Z(\sigma _0)&=&\int_{0}^{\infty }dp\,\left[
  \vphantom{\arctan\frac{p}{\frac{1}{2}+\tanh\sigma _0}}
 \coth\pi p
 \left(\arctan\frac{p}{1+\tanh\sigma _0}+3\arctan\frac{p}{\tanh\sigma _0}-\arctan\frac{p}{2}
   \right.\right.
\nonumber \\
&&\left.\left. \vphantom{\arctan\frac{p}{1+\tanh\sigma _0}}
-6\arctan p+\frac{3\pi }{2}\right)
 -4\tanh\pi p\left(\arctan\frac{p}{\frac{1}{2}+\tanh\sigma _0}-2\arctan 2p
 \right.\right.
\nonumber \\
&&\left.\left.\vphantom{\arctan\frac{p}{\frac{1}{2}+\tanh\sigma _0}}
+
 \arctan\frac{2p}{3}+\pi \right)
 +8Rp\left(\coth\pi p-\tanh\pi p\right)
 \right].
\end{eqnarray}
The easiest way to compute the integral is by differentiation in $\sigma _0$:
\begin{eqnarray}
 \frac{d}{d\sigma _0}\,\ln Z(\sigma _0)&=&\frac{1}{\cosh^2\sigma _0}\,
\int\limits_0^\infty dp\,p\left[
\frac{4\tanh  {\pi p}}{p^2+\left(\frac{1}{2}+\tanh\sigma _0\right)^2}
-\frac{\coth  {\pi p}}{p^2+\left(1+\tanh\sigma _0\right)^2}
\right.
\nonumber \\
&&\left.
\vphantom{\frac{4\tanh  {\pi p}}{p^2+\left(\frac{1}{2}+\tanh\sigma _0\right)^2}
}
-\frac{3\coth  {\pi p}}{p^2+\tanh^2\sigma _0}
\right]
+\frac{dR}{d\sigma _0}\,.
\end{eqnarray}
The following identities reduce the remaining integral to elementary functions:
\begin{eqnarray}
 &&\tanh\pi p=1-\frac{2}{\,{\rm e}\,^{2\pi p}+1}\,,\qquad 
 \coth\pi p=1+\frac{2}{\,{\rm e}\,^{2\pi p}-1}\,,
\nonumber \\
&&\int\limits_0^\infty  {\frac{dp\,p}{{\left( {{e^{2\pi p}} + 1} \right)\left( {{p^2} + {c^2}} \right)}}}  =  - \frac{{\ln c}}{2} +\frac{1}{2}\psi \left( {c + \frac{1}{2}} \right) ,
\nonumber \\
&&\int\limits_0^\infty  {\frac{dp\,p}{{\left( {{e^{2\pi p}} - 1} \right)\left( {{p^2} + {c^2}} \right)}}}  = \frac{{\ln c}}{2} - \frac{1}{{4c}} - \frac{1}{2}\psi \left( c \right),
\end{eqnarray}
and we get:
\begin{equation}
  \frac{d}{d\sigma _0}\,\ln Z(\sigma _0)=\frac{1}{2\cosh^2\sigma _0}\left(
  \frac{1}{\tanh\sigma _0+1}-\frac{3}{\tanh\sigma _0}
  \right)+\frac{dR}{d\sigma _0}.
\end{equation}

Integration over $\sigma _0$ gives:
\begin{equation}
 \Gamma _{\rm 1-loop}\equiv \ln\frac{Z(\infty )}{Z(\sigma _0)}=
 \frac{3}{2}\,\ln\tanh\sigma _0-\frac{1}{2}\,\ln\frac{\tanh\sigma _0+1}{2}+R(\infty )-R(\sigma _0).
\end{equation}
The first two terms arise from the determinants normalized by the free Klein-Gordon/Dirac operators and agree with the calculation based on the Gel'\-fand-Yaglom method \cite{Forini:2015bgo,Faraggi:2016ekd}. The last two terms is the IR anomaly. Re-expressing the coordinate cutoff  $R$ through the invariant cutoff according to (\ref{RR-inv}), we find that the last three terms cancel and we are left with 
\begin{equation}
 \Gamma _{\rm 1-loop}=\frac{3}{2}\,\ln\tanh\sigma _0=\frac{3}{2}\,\ln\cos\theta  _0,
\end{equation}
in perfect agreement with the localization prediction (\ref{localizationpredictiontheta}), (\ref{1looplocalization}).

\section{Conclusions}

The IR anomaly, related to the singular nature of the conformal gauge,  brings quantum string corrections computed in  \cite{Forini:2015bgo,Faraggi:2016ekd} in agreement with localization predictions. We also checked that the conformal anomaly cancels in each individual expectation value, even before taking the ratio. This is an important consistency check of the underlying assumptions  behind this calculation (for example, that ghosts and longitudinal modes mutually cancel in the ratio, or that the measure factors are constant and do not depend on the parameters of the problem).

One can use other parameters to build ratios of Wilson loops which are easier to compute in string theory, for instance an overall coupling to scalars as in  \cite{Beccaria:2017rbe}. But a really interesting problem is to carry out a complete calculation of quantum corrections for a single Wilson loop. A Wilson loop is a well-defined operator in field theory, and a holographic prescription to compute its expectation value in string theory should be unambiguously defined. 

\subsection*{Acknowledgements}
We would like to thank J.~Aguilera-Damia, A.~Dekel, D.~Fioravanti, Yu.~Makeenko, A.~Tseytlin and E.~Vescovi for interesting discussions. The work of D.~M.-R. and K.~Z. was supported by the ERC advanced grant No 341222. The work of K.~Z. was additionally supported by the Swedish Research Council (VR) grant
2013-4329, by the grant "Exact Results in Gauge and String Theories" from the Knut and Alice Wallenberg foundation, and by RFBR grant 15-01-99504.

\appendix

\section{Conformal anomalies}\label{conformalan-app}

Consider a second-order differential operator
\begin{equation}\label{K(alpha)}
 \mathcal{K}(\alpha )=\,{\rm e}\,^{2\alpha \phi }\left(-D_\mu D^\mu +E\right),
\end{equation}
where $D_\mu =\partial _\mu +iA_\mu $, and $A_\mu $ and $E$ are $n\times n$ matrices.
The bosonic fluctuation operators in (\ref{KGDir}), (\ref{ScaledK1})--(\ref{ScaledK3}) can be all brought to this form. The fermionic operator (\ref{f-trans}), (\ref{rescaledFermion}) has the standard Dirac form\footnote{Here we assume that the connection is Abelian, and that $A_\mu $ is a one-component $U(1)$ gauge field.}:
\begin{equation}\label{D(alpha)}
 \mathcal{D}(\alpha )=\,{\rm e}\,^{\frac{3\alpha \phi }{2}}
 \left(i\gamma ^\mu D_\mu +\gamma ^3a+v\right)\,{\rm e}\,^{-\frac{\alpha \phi }{2}},
\end{equation}
if we choose the basis of 2d gamma-matrices to be $\gamma ^\tau=-\tau _2$, $\gamma ^\sigma =\tau _1$, and $\gamma ^3\equiv  -i\varepsilon _{\mu \nu }\gamma ^\mu \gamma ^\nu /2=\tau _3$. The parameter $\alpha $ is introduced for convenience, to interpolate between tilded ($\alpha =0$) and untilded ($\alpha =1$) operators. The dependence of the determinants of $\mathcal{K}(\alpha ) $  and $\mathcal{D}(\alpha )$ on $\alpha $ is a textbook example of the  anomaly \cite{Schvarz,Fursaev-Vassilevich}. Here we give a concise derivation, that closely follows \cite{Fursaev-Vassilevich}.

The zeta-function regularized determinant of $\mathcal{K}(\alpha )$ is defined through the Mellin transform of its heat kernel:
\begin{equation}
 \ln\det\mathcal{K}=-\lim_{s\rightarrow 0}\frac{d}{ds}\,\,
 \frac{1}{\Gamma (s)}\int_{0}^{\infty }dt\,t^{s-1}\mathop{\mathrm{Tr}}\,{\rm e}\,^{-t\mathcal{K}}.
\end{equation}
Taking into account that
\begin{equation}
 \frac{\partial }{\partial \alpha }\,\mathop{\mathrm{Tr}}\,{\rm e}\,^{-t\mathcal{K}}=2t\,\frac{\partial }{\partial t}\,
 \mathop{\mathrm{Tr}}\phi \,{\rm e}\,^{-t\mathcal{K}},
\end{equation}
we find that
\begin{equation}
 \frac{d}{d\alpha }\,\ln\det\mathcal{K}
 =2\lim_{s\rightarrow 0}\frac{d}{ds}\,\,
 \frac{s}{\Gamma (s)}\int_{0}^{\infty }dt\,t^{s-1}\mathop{\mathrm{Tr}}\phi \,{\rm e}\,^{-t\mathcal{K}}.
\end{equation}
Since the gamma-function has a pole at zero, the right-hand-side seems to vanish, which would indicate that the determinant of $\mathcal{K}(\alpha )$ does not depend on the scale factor at all. But the Mellin transform generates a pole at $s=0$, because the integrand is badly behaved at $t=0$. Indeed, for any function $f(t)$ that admits a finite Laurent expansion at zero, the residue of the Mellin transform coincides with the residue of the function itself:
$$
\int_{0}^{\infty }dt\, t^{s}f(t)\stackrel{s\rightarrow 0}{=}\frac{1}{s}\,\mathop{\mathrm{res}}_{t=0}{f(t)}+{\rm regular}.
$$
The small-$t$ behaviour of the heat kernel is controlled by the DeWitt-Seeley expansion:
\begin{equation}
 \mathop{\mathrm{Tr}}\phi \,{\rm e}\,^{-t\mathcal{K}}=\sum_{k=0}^{\infty }
 t^{\frac{k}{2}-1}a_k(\phi |\mathcal{K}),
\end{equation}
where $a_k$ are local functionals of $\phi $, $E$ and $A_\mu $ that can be computed algebraically. We thus find that
\begin{equation}\label{anomaly-main}
 \frac{d}{d\alpha }\,\ln\det\mathcal{K}
 =2a_2(\phi |\mathcal{K}).
\end{equation}


The second DeWitt-Seeley coefficient of the operator (\ref{K(alpha)}) is
\begin{equation}
 a_2(\phi |\mathcal{K})=-\frac{1}{4\pi }\int_{}^{}d^2\sigma \,
 \left(\frac{\alpha n}{3}\,\partial _\mu \phi  \partial ^\mu \phi +\phi \mathop{\mathrm{tr}}E-\frac{n}{2}\,\partial _\mu \partial ^\mu \phi \right).
\end{equation}
Integrating (\ref{anomaly-main}) we express the anomaly as a local functional of the fields:
\begin{equation}\label{concrete-anomaly}
 \ln\frac{\det\mathcal{K}(1)}{\det\mathcal{K}(0)}
 =-\frac{1}{2\pi }\int_{}^{}d^2\sigma \,
 \left(\frac{n}{6}\,\partial _\mu \phi \partial ^\mu \phi +\phi \mathop{\mathrm{tr}}E-\frac{n}{2}\,\partial _\mu \partial ^\mu \phi \right).
\end{equation}

Written that way, the anomaly does not have any boundary terms\footnote{Here we assume that the metric is flat and the boundary is straight. These simplifying assumptions are sufficient for our analysis. Otherwise curvature also contributes to the anomaly.}, but it is more natural to represent it in a different form:
\begin{equation}\label{b-anomaly}
 \ln\frac{\det\mathcal{K}(1)}{\det\mathcal{K}(0)}
 =\frac{1}{2\pi }\int_{}^{}d^2\sigma \,
 \left(\frac{n}{6}\, \phi \partial _\mu\partial ^\mu \phi -\phi \mathop{\mathrm{tr}}E\right) +\frac{n}{12\pi }\oint ds\,
 \left(
 \phi \partial _n\phi - 3\partial _n\phi 
 \right),
\end{equation}
where $\partial _n$ is the inward normal derivative at the boundary. The bulk and boundary terms in the anomaly actually arise in the computation of the Seeley coefficient exactly as written in the last expression. The more compact two-dimensional form is obtained upon integration by parts.
 
To compute the anomaly for fermions we first square the Dirac operator and then by the same chain of argument that led to (\ref{anomaly-main}) arrive at
\begin{equation}\label{Dirac-anomaly-main}
 \frac{d}{d\alpha }\,\ln\det\mathcal{D}^2
 =2a_2(\phi |\mathcal{D}^2).
\end{equation}
The square of the Dirac operator (\ref{D(alpha)}) is
\begin{eqnarray}
 \mathcal{D}^2(\alpha )&=&\,{\rm e}\,^{2\alpha \phi }
 \left[
 -\nabla_\mu \nabla^\mu +\frac{\alpha }{2}\,\partial _\mu \partial ^\mu \phi 
 +a^2-v^2
 \right.
\nonumber \\
&&\left.\vphantom{\frac{\alpha }{2}}
 +\varepsilon ^{\mu \nu }\left(\partial _\mu a+\alpha a\partial _\mu \phi \right)\gamma _\nu +\left(\frac{1}{2}\,\varepsilon ^{\mu \nu }F_{\mu \nu }+2av\right)\gamma ^3
 \right],
\end{eqnarray}
where
\begin{equation}
 \nabla_\mu =D_\mu -iv\gamma _\mu -\frac{i\alpha }{2}\,\varepsilon ^{\mu \nu }\partial _\nu \phi \gamma ^3.
\end{equation}
This operator has the form (\ref{K(alpha)}) and its second DeWitt-Seeley coefficient can be read off (\ref{concrete-anomaly}):
\begin{equation}
 a_2(\phi |\mathcal{D}^2)=\frac{1}{4\pi }\int_{}^{}d^2\sigma \,\left[
 \frac{\alpha }{3}\,\partial _\mu \phi \partial ^\mu \phi +2\phi \left(v^2-a^2
 \right)
 -{\alpha }\partial _\mu \left(\phi \partial ^\mu \phi \right)
 +\partial _\mu \partial ^\mu \phi 
 \right].
\end{equation}
We thus find for the fermion anomaly:
\begin{eqnarray}\label{f-anomaly}
\frac{1}{2}\, \ln\frac{\det\mathcal{D}^2(0)}{\det\mathcal{D}^2(1)}
 &=&\frac{1}{2\pi }\int_{}^{}d^2\sigma \,
 \left[
  \frac{1 }{12}\,\phi  \partial _\mu \partial ^\mu \phi +\phi \left(a^2-v^2\right)
 \right]
\nonumber \\
&& -\frac{1}{12\pi }\oint ds\,
 \left(
 \phi \partial _n\phi - 3\partial _n\phi 
 \right).
\end{eqnarray}
 Notice that the boundary anomaly has the same magnitude but different sign compared to bosons.

\bibliographystyle{nb}

\begin{thebibliography}{10}
\ifx\href\asklfhas\newcommand{\href}[2]{#2}\fi
\raggedright
\small
\parskip 0pt

\bibitem{Forini:2015bgo}
V.~Forini, V.~Giangreco, M.~Puletti, L.~Griguolo, D.~Seminara and E.~Vescovi,
\textit{``{Precision calculation of 1/4-BPS Wilson loops in AdS$_5\times
  S^5$}''},
\textsf{JHEP~1602,~105~(2016)},
\href{http://arXiv.org/abs/1512.00841}{\texttt{1512.00841}}.
%
\bibitem{Faraggi:2016ekd}
A.~Faraggi, L.~A.~Pando~Zayas, G.~A.~Silva and D.~Trancanelli,
\textit{``{Toward precision holography with supersymmetric Wilson loops}''},
\textsf{JHEP~1604,~053~(2016)},
\href{http://arXiv.org/abs/1601.04708}{\texttt{1601.04708}}.
%
\bibitem{Forini:2017whz}
V.~Forini, A.~A.~Tseytlin and E.~Vescovi,
\textit{``{Perturbative computation of string one-loop corrections to Wilson
  loop minimal surfaces in AdS$_5 \times$ S$^5$}''},
\textsf{JHEP~1703,~003~(2017)},
\href{http://arXiv.org/abs/1702.02164}{\texttt{1702.02164}}.
%
\bibitem{Maldacena:1998im}
J.~M.~Maldacena,
\textit{``{Wilson loops in large N field theories}''},
\textsf{Phys.~Rev.~Lett.~80,~4859~(1998)},
\href{http://arXiv.org/abs/hep-th/9803002}{\texttt{hep-th/9803002}}.
%
\bibitem{Rey:1998ik}
S.-J.~Rey and J.-T.~Yee,
\textit{``{Macroscopic strings as heavy quarks in large N gauge theory and
  anti-de Sitter supergravity}''},
\textsf{Eur.~Phys.~J.~C22,~379~(2001)},
\href{http://arXiv.org/abs/hep-th/9803001}{\texttt{hep-th/9803001}}.
%
\bibitem{Drukker:1999zq}
N.~Drukker, D.~J.~Gross and H.~Ooguri,
\textit{``{Wilson loops and minimal surfaces}''},
\textsf{Phys.~Rev.~D60,~125006~(1999)},
\href{http://arXiv.org/abs/hep-th/9904191}{\texttt{hep-th/9904191}}.
%
\bibitem{Erickson:2000af}
J.~K.~Erickson, G.~W.~Semenoff and K.~Zarembo,
\textit{``{Wilson loops in N = 4 supersymmetric Yang-Mills theory}''},
\textsf{Nucl.~Phys.~B582,~155~(2000)},
\href{http://arXiv.org/abs/hep-th/0003055}{\texttt{hep-th/0003055}}.
%
\bibitem{Drukker:2000rr}
N.~Drukker and D.~J.~Gross,
\textit{``{An exact prediction of N = 4 SUSYM theory for string theory}''},
\textsf{J.~Math.~Phys.~42,~2896~(2001)},
\href{http://arXiv.org/abs/hep-th/0010274}{\texttt{hep-th/0010274}}.
%
\bibitem{Pestun:2007rz}
V.~Pestun,
\textit{``{Localization of gauge theory on a four-sphere and supersymmetric
  Wilson loops}''},
\textsf{Commun.Math.Phys.~313,~71~(2012)},
\href{http://arXiv.org/abs/0712.2824}{\texttt{0712.2824}}.
%
\bibitem{Zarembo:2016bbk}
K.~Zarembo,
\textit{``{Localization and AdS/CFT Correspondence}''},
\textsf{J.~Phys.~A50,~443011~(2017)},
\href{http://arXiv.org/abs/1608.02963}{\texttt{1608.02963}}.
%
\bibitem{Drukker:2000ep}
N.~Drukker, D.~J.~Gross and A.~A.~Tseytlin,
\textit{``{Green-Schwarz string in $AdS_5\times S^5$: Semiclassical partition
  function}''},
\textsf{JHEP~0004,~021~(2000)},
\href{http://arXiv.org/abs/hep-th/0001204}{\texttt{hep-th/0001204}}.
%
\bibitem{Kruczenski:2008zk}
M.~Kruczenski and A.~Tirziu,
\textit{``{Matching the circular Wilson loop with dual open string solution at
  1-loop in strong coupling}''},
\textsf{JHEP~0805,~064~(2008)},
\href{http://arXiv.org/abs/0803.0315}{\texttt{0803.0315}}.
%
\bibitem{Kristjansen:2012nz}
C.~Kristjansen and Y.~Makeenko,
\textit{``{More about One-Loop Effective Action of Open Superstring in
  $AdS_5\times S^5$}''},
\textsf{JHEP~1209,~053~(2012)},
\href{http://arXiv.org/abs/1206.5660}{\texttt{1206.5660}}.
%
\bibitem{Bergamin:2015vxa}
R.~Bergamin and A.~A.~Tseytlin,
\textit{``{Heat kernels on cone of $AdS_2$ and $k$-wound circular Wilson loop
  in $AdS_5 \times S^5$ superstring}''},
\textsf{J.~Phys.~A49,~14LT01~(2016)},
\href{http://arXiv.org/abs/1510.06894}{\texttt{1510.06894}}.
%
\bibitem{Drukker:2006ga}
N.~Drukker,
\textit{``{1/4 BPS circular loops, unstable world-sheet instantons and the
  matrix model}''},
\textsf{JHEP~0609,~004~(2006)},
\href{http://arXiv.org/abs/hep-th/0605151}{\texttt{hep-th/0605151}}.
%
\bibitem{Drukker:2007dw}
N.~Drukker, S.~Giombi, R.~Ricci and D.~Trancanelli,
\textit{``{More supersymmetric Wilson loops}''},
\textsf{Phys.~Rev.~D76,~107703~(2007)},
\href{http://arXiv.org/abs/0704.2237}{\texttt{0704.2237}}.
%
\bibitem{Pestun:2009nn}
V.~Pestun,
\textit{``{Localization of the four-dimensional N=4 SYM to a two-sphere and 1/8
  BPS Wilson loops}''},
\textsf{JHEP~1212,~067~(2012)},
\href{http://arXiv.org/abs/0906.0638}{\texttt{0906.0638}}.
%
\bibitem{Zarembo:2002an}
K.~Zarembo,
\textit{``{Supersymmetric Wilson loops}''},
\textsf{Nucl.~Phys.~B643,~157~(2002)},
\href{http://arXiv.org/abs/hep-th/0205160}{\texttt{hep-th/0205160}}.
%
\bibitem{Chu:2009qt}
S.-x.~Chu, D.~Hou and H.-c.~Ren,
\textit{``{The Subleading Term of the Strong Coupling Expansion of the
  Heavy-Quark Potential in a N=4 Super Yang-Mills Vacuum}''},
\textsf{JHEP~0908,~004~(2009)},
\href{http://arXiv.org/abs/0905.1874}{\texttt{0905.1874}}.
%
\bibitem{Forini:2010ek}
V.~Forini,
\textit{``{Quark-antiquark potential in AdS at one loop}''},
\textsf{JHEP~1011,~079~(2010)},
\href{http://arXiv.org/abs/1009.3939}{\texttt{1009.3939}}.
%
\bibitem{Chen-Lin:2017pay}
X.~Chen-Lin, D.~Medina-Rincon and K.~Zarembo,
\textit{``{Quantum String Test of Nonconformal Holography}''},
\textsf{JHEP~1704,~095~(2017)},
\href{http://arXiv.org/abs/1702.07954}{\texttt{1702.07954}}.
%
\bibitem{Drukker:2005cu}
N.~Drukker and B.~Fiol,
\textit{``{On the integrability of Wilson loops in $AdS_5\times S^5$: Some
  periodic ansatze}''},
\textsf{JHEP~0601,~056~(2006)},
\href{http://arXiv.org/abs/hep-th/0506058}{\texttt{hep-th/0506058}}.
%
\bibitem{Greensite:1999jw}
J.~Greensite and P.~Olesen,
\textit{``{World sheet fluctuations and the heavy quark potential in the AdS /
  CFT approach}''},
\textsf{JHEP~9904,~001~(1999)},
\href{http://arXiv.org/abs/hep-th/9901057}{\texttt{hep-th/9901057}}.
%
\bibitem{Forste:1999qn}
S.~Forste, D.~Ghoshal and S.~Theisen,
\textit{``{Stringy corrections to the Wilson loop in N=4 superYang-Mills
  theory}''},
\textsf{JHEP~9908,~013~(1999)},
\href{http://arXiv.org/abs/hep-th/9903042}{\texttt{hep-th/9903042}}.
%
\bibitem{Kinar:1999xu}
Y.~Kinar, E.~Schreiber, J.~Sonnenschein and N.~Weiss,
\textit{``{Quantum fluctuations of Wilson loops from string models}''},
\textsf{Nucl.~Phys.~B583,~76~(2000)},
\href{http://arXiv.org/abs/hep-th/9911123}{\texttt{hep-th/9911123}}.
%
\bibitem{Forini:2015mca}
V.~Forini, V.~G.~M.~Puletti, L.~Griguolo, D.~Seminara and E.~Vescovi,
\textit{``{Remarks on the geometrical properties of semiclassically quantized
  strings}''},
\textsf{J.~Phys.~A48,~475401~(2015)},
\href{http://arXiv.org/abs/1507.01883}{\texttt{1507.01883}}.
%
\bibitem{Weisberger:1987kh}
W.~I.~Weisberger,
\textit{``{Conformal Invariants for Determinants of Laplacians on Riemann
  Surfaces}''},
\textsf{Commun.~Math.~Phys.~112,~633~(1987)}.
%
\bibitem{AGD}
A.~A.~Abrikosov, L.~P.~Gorkov and I.~E.~Dzyaloshinski,
\textit{``Methods of quantum field theory in statistical physics''},
Prentice-Hall (1963).
%
\bibitem{rosen1932vibrations}
N.~Rosen and P.~M.~Morse,
\textit{``On the vibrations of polyatomic molecules''},
\textsf{Physical~Review~42,~210~(1932)}.
%
\bibitem{Beccaria:2017rbe}
M.~Beccaria, S.~Giombi and A.~Tseytlin,
\textit{``{Non-supersymmetric Wilson loop in N=4 SYM and defect 1d CFT}''},
\href{http://arXiv.org/abs/1712.06874}{\texttt{1712.06874}}.
%
\bibitem{Schvarz}
A.~S.~Schwarz,
\textit{``Quantum field theory and topology''},
Springer (1993).
%
\bibitem{Fursaev-Vassilevich}
D.~Fursaev and D.~Vassilevich,
\textit{``Operators, geometry and quanta''},
Springer (2011).
%
\end{thebibliography}

\end{document}